\documentstyle[12pt]{article}
\begin{document}

\title{Comment on ``quantum theory for mesoscopic electric circuits''
cond-mat/9907171 and cond-mat/9606206}
\author{J. C. Flores}
\date{Universidad de Tarapac\'a, Departamento de F\'\i sica, Casilla 7-D, Arica,
Chile}
\maketitle

In references con-mat/9907171 and cond-mat/9606206 (Phys.Rev.B{\bf \ 53},
4927 (1996)) by You-Quan Li and Bin Chen, was considered a mesoscopic $LC$
circuit with charge discreteness. So, it was proposed a finite difference
Schr\"odinger equation for the charge time behavior. In this comment, we
generalize the corresponding mesoscopic Hamiltonian in order to taken into
account the dissipative effects (resistance $R)$. Namely, a quantum term $R%
\frac{dq}{dt},$ proportional to the current, is added to the mesoscopic $LC$
circuit equation. This is carried-out in analogy with the theory of
Caldirola-Kanai for quantum one particle damping.

\newpage\ 

Since nanotechnology structures are in a rapid development, it seems natural
ask about the quantization for mesoscopic electric circuits. An interesting
proposition was considered in reference [1,2] for LC circuits (with
electromotive force $\varepsilon $). In fact the classical equation for the
charge evolution $d^2q/dt^2+q/LC-\varepsilon /L=0$ can be obtained from the
classical Hamiltonian

\begin{equation}
H_{clas}=\frac 1{2L}P^2+\frac 1{2C}q^2+\varepsilon q, 
\end{equation}
where $P=Ldq/dt$ is related to the current. So, the quantization of the LC
circuit seems direct [3], nevertheless, as pointed out in [1,2], the
quantization of the charge must be considered in nanostructures. It was
proposed the charge operator $\widehat{q}$ with discrete eigenvalues given by

\begin{equation}
\widehat{q}\mid n\rangle =nq_e\mid n\rangle , 
\end{equation}
where $n$ is a integer ($n\in Z$) and $q_e$ the elementary charge. So, the
discreteness of the electrical charge must be related to a finite difference
Schr\"odinger equation :

\begin{equation}
-\frac{\hbar ^2}{2q_e^2L}\left\{ \psi _{n+1}+\psi _{n-1}-2\psi _n\right\}
+\left\{ \frac{q_e^2}{2C}n^2+\varepsilon q_en\right\} \psi _n=E\psi _n, 
\end{equation}
which can be related to persistent current in mesoscopic ring, Coulomb
blockage phenomenon and others.

In the charge representation (2), the quantum Hamiltonian of the systems is
given by

\begin{equation}
\widehat{H}=-\frac{\hbar ^2}{2q_e^2L}\widehat{T}+\left\{ \frac 1{2C}\widehat{%
q}^2+\varepsilon \widehat{q}\right\} , 
\end{equation}
where the operators $\widehat{T}$ and $\widehat{q}$ are

\begin{equation}
\widehat{T}=\sum_n\left( \mid n\rangle \langle n+1\mid +\mid n+1\rangle
\langle n\mid -2\mid n\rangle \langle n\mid \right) ;\widehat{q}%
=q_e\sum_n\left( n\mid n\rangle \langle n\mid \right) .
\end{equation}

Moreover, the current operator $\widehat{I}=\frac L{i\hbar }[\widehat{H},%
\widehat{q}]$ is given by

\begin{equation}
\widehat{I}=\frac \hbar {i2q_e}\sum_n\left( \mid n\rangle \langle n+1\mid
-\mid n+1\rangle \langle n\mid \right) . 
\end{equation}

A dissipative term like $R\frac{dq}{dt}$ in the time evolution equation for
the charge, i.e. the usual dissipative term in the circuit with resistance $%
R $, can be incorporated. To do this, we consider the similarity with the
Caldirola-Kanai theory for one particle dissipation in quantum mechanics
[4-7]. Namely, consider the time depending Hamiltonian

\begin{equation}
\widehat{H}=-e^{-Rt/L}\frac{\hbar ^2}{2q_e^2L}\widehat{T}+e^{+Rt/L}\left\{
\frac 1{2C}\widehat{q}^2+\varepsilon \widehat{q}\right\} . 
\end{equation}
In fact, since $\frac d{dt}=\frac 1{ih}\left[ \widehat{H},\right] +\frac
\partial {\partial t}$ and $[\widehat{T},\widehat{I}]=0$, the motion
equation for the charge operator becomes

\begin{equation}
L\frac{d^2\widehat{q}}{dt^2}=\frac 1{2q_e^2}\left[ \left\{ \frac 1{2C}%
\widehat{q}^2+\varepsilon \widehat{q}\right\} ,\left[ \widehat{T},\widehat{q}%
\right] \right] -R\frac{d\widehat{q}}{dt}, 
\end{equation}
where the double commutator, in the right hand, corresponds to the
expression without dissipation. So, the Hamiltonian (7), related to the
evolution operator of the systems, contains dissipation. Namely, it
describes the quantum version of the LCR circuit.

A remark concerning the commutation rule for charge and current: it is
energy depending due to charge discreteness and given by [1,2]:

$$
[\widehat{q},\widehat{I}]=i\hbar \left( 1+\frac{q_e^2L}{\hbar ^2}\widehat{H}%
_o\right) . 
$$
Similar generalized commutation rules were considered in high energy physics
with space quantization [8,9]. Namely, for coordinate and momentum, it was
considered $[x,p]=i\hbar \left( 1+sH_o\right) ,$ where $s$ is a parameter
related to the space discreteness.

A final comment respect to the electromotive force incorporated in (3): in
this theory, $\varepsilon $ is a quantum (microscopic) electromotive force,
but in usual nanoestructures circuit this source is macroscopic i.e. it must
be considered as producing decoherence and mixing.

$$
{} 
$$

Acknowledgments: a preliminary discussion for quantum capacitances was
carried-out with Professor P. Orellana (UCN). This work was partially
supported by grant FONDECYT 1990443.

\end{document}